# Infrared band strengths of dangling OH features in amorphous water at 20 K

Takeshi Hasegawa,[1] Hiroto Yanagisawa,[1,2] Takumi Nagasawa,[1,3] Reo Sato[1], Naoki Numadate[1,4], and Tetsuya Hama[1,5]

[1] Komaba Institute for Science and Department of Basic Science, The University of Tokyo, Meguro, Tokyo 153-8902, Japan.

[2] Present affiliations: Institute for Cosmic Ray Research, The University of Tokyo, 5-1-5 Kashiwanoha, Kashiwa, Chiba 277-8582, Japan. Department of Physics, Graduate School of Science, The University of Tokyo, 7-3-1 Hongo, Bunkyo, Tokyo 113-0033, Japan.

[3] Present affiliations: Komaba Institute for Science and Department of Basic Science, The University of Tokyo, Meguro, Tokyo 153-8902, Japan. Univ. Grenoble-Alpes, CNRS, LIPhy, 38000 Grenoble, France.

[4] Present affiliation: Department of Chemistry, Faculty of Pure and Applied Sciences, University of Tsukuba, Tsukuba 305-8571, Japan.

[5]Author to whom correspondence should be addressed: hamatetsuya@g.ecc.u-tokyo.ac.jp

Short title: Band strengths of dangling OH features

**Abstract**

Infrared (IR) spectra of vapor-deposited amorphous water at low temperatures show two weak peaks at around 3720 and 3696 cm$^{-1}$ assigned to free-OH stretching modes of two- and three-coordinated water molecules (so-called 'dangling' OH bonds), respectively, on the ice surface. A recent James Webb Space Telescope (JWST) observation first succeeded in detection of a potential dangling OH feature at 3664 cm$^{-1}$ for ices in molecular clouds, highlighting the importance of dangling OH bonds in interstellar ice chemistry. A lack of band strengths of these features at low temperatures restricts the quantification of dangling OH bonds from IR spectra, hindering development of a molecular-level understanding of the surface structure and chemistry of ice. Using IR multiple-angle incidence resolution spectrometry, we quantified the band strengths of two- and three-coordinated dangling OH features in amorphous water at 20 K as being 4.6 ± 1.6 × 10$^{-18}$ and 9.1 ± 1.0 × 10$^{-18}$ cm molecule$^{-1}$, respectively. These values are more than an order of magnitude lower than band strengths of bulk water molecules in ice and liquid water and are similar to those of H$_2$O monomers confined in solid matrices. Adsorption of carbon monoxide with dangling OH bonds results in the appearance of a new broad dangling OH feature at 3680–3620 cm$^{-1}$, with a band strength of 1.8 ± 0.1 × 10$^{-17}$ cm molecule$^{-1}$. The band strengths of dangling OH features determined in this study advance our understanding of the surface structure of interstellar ice analogs and recent IR observations of the JWST.

# 1. Introduction

In cold interstellar regions, such as molecular clouds, dust is covered by molecular solids comprising mainly water ice (Van Dishoeck et al. 2013; Hama & Watanabe 2013). These icy dust grains are precursors of planetary material. Interstellar ices can be detected by their infrared (IR) absorption at around 3300 cm$^{-1}$ (3.0 μm) (Boogert et al. 2015). This absorption band originates from the OH stretching vibrational mode of bulk $H_2O$ molecules with four-coordinated hydrogen-bonded structures in the ice; each water molecule donates hydrogen bonds to two neighbors and accepts bonds from two other neighbors (Bartels-Rausch et al. 2012). The bulk OH stretching band observed in molecular clouds is broad (with no sharp peak) at around 3200 cm$^{-1}$ (3.1 μm), which indicates that ices in molecular clouds are in metastable amorphous forms (Boogert et al. 2015; Van Dishoeck et al. 2013; Hama & Watanabe 2013; McClure et al. 2023).

Laboratory studies have shown that IR spectra of porous amorphous water exhibit two weak features at 3720 and 3696 cm$^{-1}$, in addition to the bulk OH stretching band (Buch & Devlin 1991; He et al. 2019; Maté et al. 2021; Noble et al. 2014a, 2014b). These features are assigned to the free-OH stretching vibrational modes of two-coordinated (one-donor and one-acceptor) and three-coordinated (one-donor and two-acceptors) water molecules at the ice surface, respectively, and are often termed 'dangling' OH bonds (Buch & Devlin 1991; Cholette et al. 2009; Devlin & Buch 1995; Nagata et al. 2019; Noble et al. 2014b, 2014a). Here, two- and three-coordinated dangling OH bonds are called 2dOH and 3dOH, respectively.

Dangling OH bonds reflect the surface structure and porosity of amorphous water, and they act as catalytic sites for adsorption and chemical reactions. For example, the 3dOH feature at 3696 cm$^{-1}$ is observed for both porous and nonporous amorphous water, whereas the 2dOH feature at 3720 cm$^{-1}$ appears only for porous amorphous water (Bu et al. 2015; He et al. 2018; Nagasawa et al. 2022, 2021). Thus, the 2dOH/3dOH intensity ratio can be used to characterize the porosity of amorphous water. In addition, various molecules, such as carbon monoxide (CO), carbon dioxide ($CO_2$), ammonia ($NH_3$), methanol ($CH_3OH$), and polycyclic aromatic hydrocarbons (PAHs), preferentially adsorb to dangling OH bonds on the ice surface (Bahr et al. 2008; He et al. 2022a, 2019; Noble et al. 2020). Unique photochemical reactions of CO and PAHs are induced when these molecules interact with dangling OH bonds (Matsuda et al. 2018; Noble et al. 2020). A recent observation by the James Webb Space Telescope (JWST) tentatively identified a dangling OH feature at 3664 cm$^{-1}$ (2.73 μm) for ices in pristine molecular clouds (McClure et al. 2023). The decrease in the peak wavenumber compared with the 2dOH and 3dOH features suggests that the dangling OH bonds in interstellar ices are interacted with other volatile molecules (e.g., CO, methane ($CH_4$), and nitrogen ($N_2$)) on the ice surface (He et al. 2018; Palumbo et al. 2010).

For quantitative understanding of the surface physics and chemistry of interstellar ices and their laboratory analogs, the 2dOH and 3dOH band strengths (integrated absorption cross-sections) are required for deriving their column densities (molecules cm$^{-2}$). The optical constants of dangling OH bonds were previously studied using IR reflection–absorption spectroscopy with metallic substrates (Cholette et al. 2009).

However, the band strengths of the 2dOH and 3dOH features are still poorly understood, and their measurement is difficult due to their intrinsic surface origins. The Beer–Lambert law is often used to describe the relationship between sample absorbance ($A$) and absorption cross-section ($\sigma$, cm$^2$ molecules$^{-1}$) at a given wavelength ($\lambda$, μm) as a function of the imaginary part ($k$) of its complex refractive index ($n + ik$), also known as the extinction coefficient (Hasegawa 2017).

$$A = -\log_{10}\frac{S^s}{S^b} = \frac{\sigma N}{\ln 10} = \frac{4\pi dk}{\lambda \ln 10} \quad (1)$$

Here, $S$ is the intensity of transmitted IR light of wavelength $\lambda$; superscripts $s$ and $b$ indicate sample and background measurements, respectively; and $d$ and $N$ represent sample thickness (μm) and column density of molecules in a sample, respectively. Hence, the absorption cross-section, $\sigma$, can be written as

$$\sigma = \frac{4\pi dk}{N\lambda} \quad (2)$$

Band strength $\beta$ (cm molecule$^{-1}$) can be obtained by integrating $\sigma$ over the wavelength region of interest. Equation (2) shows that values for both $k$ and $N$ of 2dOH and 3dOH are required to obtain their $\sigma$ and $\beta$ values, by definition. However, 2dOH and 3dOH are located on the ice surface (including the internal pore surface in the bulk ice), so absorbance induced by them is not necessarily proportional to sample thickness (Nagasawa et al. 2021). This means that application of the Beer–Lambert law (Equation (1)) is not straightforward for 2dOH and 3dOH (Nagasawa et al. 2021, 2022). Indeed, it holds only for a bulk sample whose thickness is greater than the IR wavelength ($d \geq \lambda$), because optical interfaces (e.g.,

vacuum–sample–substrate interfaces) are not considered in the derivation of the law from Maxwell equations (Hasegawa 2017).

To overcome this problem, we developed an experimental setup with IR multiple-angle incidence resolution spectrometry (IR–MAIRS) under low-temperature and ultrahigh-vacuum conditions (see Appendix for details). IR–MAIRS is a spectroscopic technique that combines oblique-incidence transmission measurements and chemometrics (multivariate analysis) to retrieve both pure in-plane (IP) and out-of-plane (OP) vibrational spectra for thin samples (Hasegawa & Shioya 2020; Shioya et al. 2019) (Fig. 1A). Although the technique was originally intended for analysis of the molecular orientation of organic thin films at atmospheric pressure, it is also useful for *in situ* structural characterization of vapor-deposited thin samples in a vacuum (Hama et al. 2020; Hasegawa & Shioya 2020; Nagasawa et al. 2022). As described in Section 2, IP and OP vibrational spectra are expressed using the surface-parallel (*x*-, and *y*-direction) and perpendicular (*z*-direction) components of the extinction coefficient of the thin sample ($k_{xy}$ and $k_z$, respectively. See Fig. 1A for the definition of the *x*-, *y*-, and *z*-directions). Because *k* is an optically isotropic coefficient, it is related to $k_{xy}$ and $k_z$ by Equation (3) (Nagasawa et al. 2022).

$$k = \frac{2k_{xy} + k_z}{3} \quad (3)$$

Consequently, *k* values for dangling OH bonds can be obtained by measuring both their IP ($k_{xy}$) and OP ($k_z$) vibrational spectra. This unique characteristic of IR–MAIRS has major advantages over conventional measurement techniques, such as normal-incidence transmission and reflection–absorption spectroscopy with metallic substrates, which observe only IP ($k_{xy}$) and OP ($k_z$) vibrations, respectively (Fig. 1B and C, see also the Appendix for details). In our previous low-temperature, ultrahigh-vacuum IR–MAIRS studies, we measured the band strength of 3dOH in amorphous water at 90 K as $1.4 \pm 0.3 \times 10^{-17}$ cm molecule$^{-1}$ at 3710–3680 cm$^{-1}$ (Nagasawa et al. 2021, 2022). Here, we report the band strengths of 2dOH and 3dOH features in amorphous water at 20 K, a temperature relevant to molecular clouds. We also derived the band strength of the broad and red-shifted dangling OH feature at 3680–3620 cm$^{-1}$ caused by adsorption of CO molecules.

## 2. Results and Discussion

### 2.1. IP and OP spectra of vapor-deposited amorphous water at 20 K

IR–MAIRS IP and OP spectra at 4000–800 cm$^{-1}$ are shown in Fig. 2A for vapor-deposited amorphous water on a Si substrate at 20 K, prepared by 32 min exposure of water at $5.6 \pm 2.0 \times 10^{-6}$ Pa, corresponding to $3.9 \pm 1.4 \times 10^{16}$ molecules cm$^{-2}$. Amorphous water samples were prepared using H$_2$O containing 3.5 mol.% HDO to allow measurement of the OD stretching vibration decoupled from intra- and inter-molecular OH stretching vibrations. There was a weak decoupled OD stretching vibrational peak at around 2400 cm$^{-1}$, in addition to OH stretching, combination, bending, and vibration bands (Eisenberg & Kauzmann 1969). The peak wavenumber of the decoupled OD stretching band is sensitive to the local lattice structure (oxygen–oxygen distance) in bulk ice (Franks 1972; Hama et al. 2017; Klug et al. 1987). Molecular orientation analysis also requires the decoupled OD stretching band because its transition moment direction reflects the molecular orientation. In contrast, bulk OH stretching vibrations in $H_2O$ ice are delocalized by intermolecular vibrational coupling, and thus are inappropriate for molecular orientation analysis (Nagasawa et al. 2022). A more detailed investigation of the oxygen–oxygen distance and molecular orientation in bulk amorphous water is currently in progress using the decoupled-OD stretching band, and will be reported separately. Here, we focus on the band strengths of 2dOH and 3dOH features (Fig. 2B), with the contribution of surface HDO molecules to dangling OH bonds being negligible due to its low concentration of 3.5 mol.%.

The 2dOH and 3dOH features appeared in both the IP and OP spectra (Fig. 2B), and because they overlap we used two Gaussian functions to fit them (Fig. 2C and D). Similar band shapes and intensities were obtained for both 2dOH and 3dOH features in IP and OP spectra (Table A1 in the Appendix). This indicates the isotropic nature (with random orientation) of 2dOH and 3dOH in amorphous water at 20 K, which is characteristic of the formation of porous amorphous water with a large internal surface area (He et al. 2019; Kimmel et al. 2001; Nagasawa et al. 2021, 2022; Stevenson et al. 1999).

Analytical expressions for absorbances in IR–MAIRS IP and OP spectra have been formulated previously, based on Maxwell's equations (Hasegawa 2017; Itoh et al. 2009; Nagasawa et al. 2022; Shioya et al. 2019). Complex mathematical calculations are required, so only the final outcomes are described here. Absorbance in an IP spectrum ($A_{IP}$) corresponds quantitatively to that of a thin sample ($d \ll \lambda$) with normal-incidence transmission using linearly polarized IR light ($A_{thin}^{\theta=0°}$) in a five-layer symmetric double-sided sample system (vacuum/thin sample/IR-transparent substrate/thin sample/vacuum), which is typical of vapor deposition (Fig. 1A and B).

$$A_{\text{IP}} = A_{\text{thin}}^{\theta=0} = \frac{8\pi da}{\lambda \ln 10} f_{\text{TO}} \quad (4)$$

$$a = \frac{1}{n_v + n_s} + \left(\frac{n_v - n_s}{n_v + n_s}\right)^4 \left\{1 - \left(\frac{n_v - n_s}{n_v + n_s}\right)^4\right\}^{-1} \left(\frac{2n_s}{n_s^2 - n_v^2}\right) \quad (5)$$

$$f_{\text{TO}} = \text{Im}[(n_{xy} + ik_{xy})^2] = 2n_{xy}k_{xy} \quad (6)$$

Here, $n_v$ and $n_s$ are the refractive index of the vacuum ($n_v = 1$) and substrate ($n_s = 3.41$ for Si), respectively (Tasumi 2014). Coefficient $a$ (0.290 for Si) expressed in terms of $n_v$ and $n_s$ is added in $A_{\text{IP}}$ and also in $A_{\text{thin}}^{\theta=0°}$ because thin-sample absorbance is induced by the electric field at optical interfaces (vacuum–sample–substrate interfaces) when sample thickness $d$ is much less than the IR wavelength $\lambda$, that is, $d \ll \lambda$ (Hasegawa 2017). The transverse optic (TO) energy-loss function of the thin sample, $f_{\text{TO}}$, is expressed using the surface-parallel (x- and y-direction) components of the complex refractive index ($n_{xy} + ik_{xy}$) of the thin sample. Consequently, $A_{\text{IP}}$ and $A_{\text{thin}}^{\theta=0°}$ include only information about the surface-parallel component of a transition moment (IP molecular vibration) of the thin sample and their band shapes are given by $f_{\text{TO}}$ (Hasegawa 2017; Itoh et al. 2009).

Absorbance in an OP spectrum ($A_{\text{OP}}$) is expressed similarly to $A_{\text{IP}}$, with the TO energy-loss function replaced by the longitudinal optic (LO) energy-loss function, $f_{\text{LO}}$, expressed using the surface-perpendicular (z-direction) component of the complex refractive index ($n_z + ik_z$) of the thin sample, as

$$A_{\text{OP}} = \frac{8\pi da}{H\lambda \ln 10} f_{\text{LO}} \quad (7)$$

$$f_{\text{LO}} = \text{Im}\left[\frac{1}{(n_z + ik_z)^2}\right] = \frac{2n_z k_z}{(n_z^2 + k_z^2)^2} \quad (8)$$

where $H$ is a substrate-specific correction factor (0.334 for Si) accounting for the intensity ratio of electric fields at the optical interface along the surface-parallel and surface-perpendicular directions (Nagasawa et al. 2022; Shioya et al. 2019).

Vapor-deposited amorphous water has values of $k$ that should be approximately two orders of magnitude lower than $n$ ($k \ll n$) at 3750–3650 cm$^{-1}$ in bulk water (Mastrapa et al. 2009), satisfying the following two approximations (Equations (9) and (10)).

$$\frac{k^2}{n^2} \ll 1 \quad (9)$$

$$n_{xy} = n_z = n \quad (10)$$

Equation (10) indicates that the thin sample has no optical anisotropy and no anomalous dispersion, consistent with the first approximation (Nagasawa et al. 2022; Shioya et al. 2017). Here we used $n = 1.20$ or $1.26$ for vapor-deposited amorphous water at 20 or 90 K, respectively (Kofman et al. 2019). In such cases, $A_{IP}$ and $A_{OP}$ were simplified to Equations (11) and (12), respectively.

$$A_{IP} \approx \frac{8\pi da}{\lambda \ln 10}(2nk_{xy}) \quad (11)$$

$$A_{OP} \approx \frac{8\pi da}{H\lambda \ln 10}\frac{(2nk_z)}{n^4} \quad (12)$$

Thus, $k_{xy}$, $k_z$, and optically isotropic $k$ can be obtained from $A_{IP}$ and $A_{OP}$, even for thin samples, as

$$k_{xy} \approx \frac{A_{IP}\lambda \ln 10}{16\pi dan} \quad (13)$$

$$k_z \approx \frac{n^4 H A_{\text{OP}} \lambda \ln 10}{16\pi dan} \quad (14)$$

$$k = \frac{2k_{xy} + k_z}{3} \approx \left(\frac{2A_{\text{IP}} + n^4 H A_{\text{OP}}}{3}\right) \frac{\lambda \ln 10}{16\pi dan} \quad (15)$$

Therefore, values of $\sigma$ and band strength ($\beta$) for 2dOH and 3dOH features can be obtained from the IP and OP spectra, as shown in Equations (16) and (17).

$$\sigma = \frac{4\pi dk}{\lambda N} \approx \frac{4\pi d}{\lambda N}\left(\frac{2A_{\text{IP}} + n^4 H A_{\text{OP}}}{3}\right)\frac{\lambda \ln 10}{16\pi dan} = \left(\frac{2A_{\text{IP}} + n^4 H A_{\text{OP}}}{3}\right)\frac{\ln 10}{4anN} \quad (16)$$

$$\beta = \int \sigma(\tilde{\nu})d\tilde{\nu} \approx \left(\frac{2}{3}\int A_{\text{IP}}(\tilde{\nu})d\tilde{\nu} + \frac{n^4 H}{3}\int A_{\text{OP}}(\tilde{\nu})d\tilde{\nu}\right)\frac{\ln 10}{4naN} \quad (17)$$

Here, $\sigma(\tilde{\nu})$ is the absorption cross-section at a given wavenumber ($\tilde{\nu}$).

Equation (17) indicates that column densities of 2dOH ($N_{\text{2dOH}}$) and 3dOH ($N_{\text{3dOH}}$) are required for calculation of their $\beta$ values ($\beta_{\text{2dOH}}$ and $\beta_{\text{3dOH}}$, respectively). The $N_{\text{3dOH}}$ value for nonporous amorphous water at 90 K has previously been estimated with methanol ($CH_3OH$) deposition because the hydrogen bonding of $CH_3OH$ with dangling OH bonds quenches 3dOH features (Nagasawa et al. 2021, 2022). However, at 20 K, $CH_3OH$ does not efficiently diffuse to 2dOH and 3dOH bonds on porous amorphous water, so cannot be used as a probe molecule to estimate $N_{\text{2dOH}}$ and $N_{\text{3dOH}}$. Therefore, we used CO deposition to estimate the column density of dangling OH bonds at 20 K for the following three reasons. First, CO can diffuse across the ice surface to find and interact with dangling OH bonds even at 20 K (Collings et al. 2005; He et al. 2018; Lauck et al. 2015). Second, the interaction between CO and dangling OH bonds quenches 2dOH and 3dOH features (He et al. 2019). Third, CO is IR-active and its band strength has been experimentally determined in previous studies (Bouilloud et al. 2015; Gerakines et al. 1995, 2023; González Díaz et al. 2022; Jiang et al. 1975; Sandford et al. 1988).

**2.2. CO deposition on porous amorphous water at 20 K**

The disappearance of 2dOH and 3dOH features in porous amorphous water at 20 K following CO deposition is illustrated in Fig. 3A. From repeated experiments, we found that 2dOH and 3dOH features completely vanished after CO deposition for 40 min at 7.1 ± 0.6 × $10^{-7}$ Pa, corresponding to exposure of 4.9 ± 0.4 × $10^{15}$ molecules $cm^{-2}$. After CO deposition, a new broad peak appeared in a lower wavenumber region at around 3650 $cm^{-1}$ (Fig. 3B and C) due to the interaction of 2dOH and 3dOH with CO (He et al. 2018, 2019). Here, this peak is termed the 'perturbed-dOH' feature.

After CO deposition, the CO stretching band appeared at around 2151 and 2139 $cm^{-1}$ in both the IP and OP spectra (Fig. 3D), which were blue- and red-shifted, respectively, from their gas-phase value (2143 $cm^{-1}$) (Mina-Camilde et al. 1996). The blue-shifted feature at 2151 $cm^{-1}$ was assigned to CO molecules hydrogen-bonded to dangling OH through the carbon atom (Al-Halabi et al. 2004; Brookes & McKellar 1998; Fraser et al. 2004; Lundell & Räsänen 1995; Ryazantsev et al. 2016; Zamirri et al. 2018). The red-shifted feature at around 2139 $cm^{-1}$ may have resulted from the CO molecules adsorbed on the surface of amorphous water through non-hydrogen-bond interactions (Zamirri et al. 2018). The 2139 $cm^{-1}$ feature may be associated with CO interacting with 'bonded-OH' water molecules, which had OH groups hydrogen-bonded with other water molecules, pointing downward on the ice surface (Al-Halabi et al. 2004). Gaussian fitting of the two blue- and red-shifted features indicated full width at half-maximum (FWHM) values larger than that of crystalline CO (1.5 cm$^{-1}$) (Fig. 3E and F) (Gerakines et al. 2023); thus, there is only a sub-monolayer coverage of CO on the ice surface. Here, the CO molecules giving the blue- and red-shifted features are termed 'dOH-CO' and 'other-CO', respectively. Further discussion of the dOH-CO and other-CO features is presented in the Appendix.

We calculated the column density of CO molecules adsorbed on the amorphous water surface from the CO stretching band area in IP and OP spectra by rearranging Equation (17) as follows.

$$N = \frac{\ln 10}{4na\beta}\left(\frac{2}{3}\int A_{\text{IP}}(\tilde{v})d\tilde{v} + \frac{n^4 H}{3}\int A_{\text{OP}}(\tilde{v})d\tilde{v}\right) \quad (18)$$

Here, $A_{\text{IP}}(\tilde{v})$ and $A_{\text{OP}}(\tilde{v})$ are IP and OP absorbance at a given wavenumber ($\tilde{v}$), respectively. It has been shown that the band strength of crystalline CO ($9.8 \times 10^{-18}$ cm molecule$^{-1}$) is a reasonable estimate of the band strength of pure amorphous CO (Gerakines et al. 2023). The CO band strength in a solid $H_2O$ + CO mixture has been experimentally determined as $1.1$–$1.7 \times 10^{-17}$ cm molecule$^{-1}$ (Gerakines et al. 1995; Sandford et al. 1988). We used the average value ($1.3 \pm 0.3 \times 10^{-17}$ cm molecule$^{-1}$) of the above published CO band strengths ($9.8 \times 10^{-18}$, $1.1 \times 10^{-17}$, and $1.7 \times 10^{-17}$ cm molecule$^{-1}$) in estimating the column density of CO molecules on the amorphous water surface.

We repeated independent statistical experiments five times and calculated the total column density of CO molecules on the amorphous water surface as $N_{\text{CO}} = 5.8 \pm 0.1 \times 10^{15}$ molecules based on band areas of Gaussian fitting for the dOH-CO and other-CO features in the IP and OP spectra at 2170–2120 cm$^{-1}$ (Fig. 3E and F, and Table A2 in the Appendix). This value corresponds to the total CO column density adsorbed on both sides of the Si substrate of sample thickness 2$d$ (Fig. 1A). Thus, there are $N_{CO}/2 = 2.9 \pm 0.2 \times 10^{15}$ molecules cm$^{-2}$ on each side of the substrate. This value is smaller than that (4.9 ± 0.4 × 10$^{15}$ molecules cm$^{-2}$) estimated from the CO exposure at $7.1 \times 10^{-7}$ Pa for 40 min. This difference can arise from a pressure gradient between the cold cathode gauge and the helium cryostat connected with the Si substrate in the vacuum chamber (Nagasawa et al. 2022). Because the helium cryostat efficiently works as a cryopump, the partial pressure of CO around the Si substate can be decreased compared with that around the cold cathode gauge.

Considering the fraction of the band area of Gaussian fitting for the dOH-CO feature (Fig. 3E and F, and Table A2 in the Appendix), the column density of dOH-CO molecules was estimated as $N_{dOH-CO} = 1.9 \pm 0.2 \times 10^{15}$ molecules cm$^{-2}$ ($N_{dOH-CO}/2 = 9.3 \pm 1.0 \times 10^{14}$ molecules cm$^{-2}$ on each side of the substrate). Here, the value of $N_{dOH-CO}$ includes both the column densities of CO molecules hydrogen-bonded with 2dOH and 3dOH, that is, $N_{dOH-CO} = N_{2dOH} + N_{3dOH}$. Therefore, it was not possible to quantify $N_{2dOH}$ and $N_{3dOH}$ directly from $N_{dOH-CO}$ values obtained from the blue-shifted dOH-CO feature. We did not decompose the blue-shifted dOH-CO feature further by Gaussian fitting. Instead, we performed the following separate experiments to obtain the band strength of the 3dOH feature ($\beta_{3dOH}$) at 20 K. Once $\beta_{3dOH}$ was obtained, $N_{3dOH}$ for porous amorphous water at 20 K could also be derived from 3dOH features in the IP and OP spectra before CO deposition (Fig. 3A(a)), eventually leading to $N_{2dOH} = N_{dOH-CO} - N_{3dOH}$.

## 2.3. Band strength of 3dOH at 20 K

To derive $\beta_{3dOH}$ at 20 K, we first prepared nonporous amorphous water by vapor deposition at 90 K (Kimmel et al. 2001; Stevenson et al. 1999). Only the 3dOH feature appeared in the OP spectrum, without the 2dOH feature (Fig. 4A(a)), and there were no clear 3dOH or 2dOH features in the IP spectrum. As discussed in our previous studies (Nagasawa et al. 2021, 2022), these results indicate that only 3dOH bonds were present on the top surface of nonporous amorphous water at 90 K, with a strong perpendicular orientation. The nonporous amorphous water was cooled to 20 K (Fig. 4A(b)), and CO was deposited on it to determine the column density of 3dOH.

After CO deposition for 12 min at $1.6 \pm 0.1 \times 10^{-7}$ Pa, corresponding to an exposure of $3.3 \pm 0.3 \times 10^{14}$ molecules cm$^{-2}$, the 3dOH feature disappeared (Fig. 4A(c)) and the CO stretching band appeared in both the IP and OP spectra (Fig. 4B(c)). Gaussian fitting of the CO stretching bands revealed that the peak positions of the dOH-CO features differed by 5.6 cm$^{-1}$ between the IP (2148.7 ± 1.0 cm$^{-1}$) and OP (2154.3 ± 0.4 cm$^{-1}$) spectra (Fig. 4C and D). The Gaussian fitting also revealed that the other-CO features had similar peak positions (2139.4 ± 0.3 cm$^{-1}$ vs. 2139.9 ± 0.1 cm$^{-1}$) and FWHM values (8.5 ± 0.4 cm$^{-1}$ vs. 8.8 ± 0.5 cm$^{-1}$) between the IP and OP spectra. The difference in dOH-CO peak positions between the spectra suggests there are at least two energetically different configurations for dOH-CO molecules (dOH-CO molecules are CO molecules adsorbed with 3dOH bonds), as shown schematically in Fig. 4E.

Previous matrix-isolation spectroscopy studies showed that the blue-shifted dOH-CO feature of the H$_2$O···CO complex in an Ar matrix has peaks at 2152, 2149, and 2148 cm$^{-1}$ depending on trapping sites, and that the higher-wavenumber peak corresponds to more energetically stable sites (Lundell & Räsänen 1995; Ryazantsev et al. 2016). Thus, the results in Fig. 4 imply that the stable 2154.3 cm$^{-1}$ configuration has a perpendicular CO orientation (OP vibration), and that the metastable 2148.7 cm$^{-1}$ configuration has a parallel CO orientation (IP vibration) relative to the ice surface. Because 3dOH bonds on nonporous amorphous water have a strong perpendicular orientation to the ice surface (Fig. 4A), the orientation of the dOH-CO molecules may also be anisotropic depending on their configurations.

Supporting this, Collings et al. (2005) acquired IR spectra of CO adsorbed on vapor-deposited water ices at sub-monolayer coverage (1 Langmuir exposure) using IR reflection–absorption spectroscopy. They prepared nonporous amorphous water on gold at 110 K and cooled it to 30 K for CO deposition, where adsorbed CO molecules were mobile and preferentially occupied the most energetically favorable adsorption sites. Accordingly, the dOH-CO peak was blue-shifted at 2156 cm$^{-1}$ (IR reflection–absorption spectroscopy with a metallic substrate selectively measures OP ($k_z$) vibration) (Fig. 1C) (Hasegawa 2017; Hasegawa & Shioya 2020; Nagasawa et al. 2022; Tolstoy et al. 2003; Yamamoto & Ishida 1994). Thus, these reported results suggest that the energetically stable configuration for dOH-CO at 2156 cm$^{-1}$ has an OP vibration perpendicular to the ice surface, consistent with our present results (Fig. 4E).

Using the band areas of Gaussian fitting for the dOH-CO feature in the IP and OP spectra at 2170–2120 cm$^{-1}$ (Fig. 4C and D), we calculated the column density of 3dOH to be $1.6 \pm 0.2 \times 10^{14}$ molecules cm$^{-2}$ ($8.2 \pm 0.9 \times 10^{13}$ molecules cm$^{-2}$ on each side of the Si substrate) for nonporous amorphous water at 20 K (Tables A3 and A4 in the Appendix). The band strength of the 3dOH feature was obtained as $\beta_{3dOH} = 9.1 \pm 1.0 \times 10^{-18}$ cm molecule$^{-1}$ at 20 K (Equation 17), using the IP and OP band areas of the 3dOH feature for the nonporous amorphous water at 20 K before CO deposition (Fig. 4A).

Our band-strength value for 3dOH at 20 K ($9.1 \pm 1.0 \times 10^{-18}$ cm molecule$^{-1}$) is slightly lower than that at 90 K ($1.4 \pm 0.3 \times 10^{-17}$ cm molecule$^{-1}$), as reported previously (Nagasawa et al. 2021, 2022). The difference in band strength between 20 and 90 K is consistent with the OP band area of the 3dOH feature for nonporous amorphous water becoming $1.4 \pm 0.2$ times smaller on cooling to 20 K from the preparation temperature of 90 K (Fig. 4A(a) and (b), and Table A5 in the Appendix), although such agreement may be fortuitous considering the systematic errors in the IR–MAIRS method. For example, estimation of the column density of dangling OH bonds relies on the band strength of the probe molecule, and we used different probe molecules at 20 (CO) and 90 K (CH$_3$OH). As described in the section 2.2, we used the average value ($1.3 \pm 0.3 \times 10^{-17}$ cm molecule$^{-1}$) of the previously published CO band strengths for the band strength of the dOH-CO feature [$9.8 \times 10^{-18}$ cm molecule$^{-1}$ for crystalline CO and $1.1 \times 10^{-17}$ and $1.7 \times 10^{-17}$ cm molecule$^{-1}$ for CO embedded in bulk H$_2$O ice (Gerakines et al. 2023, 1995; Sandford et al. 1988)]. If the band strength of the dOH-CO feature is largely different from the above literature values, the band strengths of dangling OH features obtained in this study should be modified. In addition, the adsorption of residual gases, such as $N_2$ on 3dOH in the vacuum chamber, may occur following the cooling of nonporous amorphous water to 20 from 90 K, reducing the band area of the 3dOH feature. Further experiments using different probe molecules (e.g., $NH_3$, $CO_2$, and $CH_4$) may improve the accuracy of band strength determinations for dangling OH bonds, but this is beyond the scope of this Letter. Nevertheless, our previous and present studies suggest little temperature dependence (within the factor of two) for the band strength of 3dOH between 20 and 90 K.

For reference, the peak position of dOH-CO molecules in porous amorphous water at 20 K was centered at 2151 cm$^{-1}$ in both the IP and OP spectra, without clear peaks at 2148.7 and 2154.3 cm$^{-1}$ (Fig. 3E and F). In the case of porous amorphous water, both 2dOH and 3dOH bonds were located not only on the top surface but also in the internal pore surface. Therefore, they have isotropic (random) orientation (Fig. 2 and 3A(a)), with molecular orientation of dOH-CO molecules also being randomized. The 2148.7 and 2154.3 cm$^{-1}$ components were averaged to yield a peak at around 2151 cm$^{-1}$ in both IP and OP spectra (Fig. 3E and F).

### 2.4. Band strengths of 2dOH and perturbed dOH at 20 K

Based on the 20 K $\beta_{3dOH}$ value (9.1 ± 1.0 × 10$^{-18}$ cm molecule$^{-1}$), the $N_{3dOH}$ value of porous amorphous water at 20 K was calculated from 3dOH features in IP and OP spectra before CO deposition (Fig. 2C, D, and 3A(a)) as $N_{3dOH}$ = 1.2 ± 0.1 × 10$^{15}$ molecules cm$^{-2}$ (6.1 ± 0.3 × 10$^{14}$ molecules cm$^{-2}$ on each side of the Si substrate, see Table A1 in the Appendix). The $N_{2dOH}$ value was also derived: $N_{2dOH} = N_{dOH-CO} - N_{3dOH} = 6.5 \pm 2.2 \times 10^{14}$ molecules cm$^{-2}$ (3.2 ± 1.1 × 10$^{14}$ molecules cm$^{-2}$ on each side); and the 2dOH peak band strength was calculated as $\beta_{2dOH} = 4.6 \pm 1.6 \times 10^{-18}$ cm molecule$^{-1}$ at 20 K.

We also calculated the band strength of the perturbed dOH feature at 20 K as $\beta_{perturbed\text{-}dOH} = 1.8 \pm 0.1 \times 10^{-17}$ cm molecule$^{-1}$ for the wavenumber region 3680–3620 cm$^{-1}$, based on the column density of perturbed dOH bonds corresponding to $N_{2dOH} + N_{3dOH} = 1.9 \pm 0.2 \times 10^{15}$ molecules cm$^{-2}$ (9.3 ± 1.0 × 10$^{14}$ molecules cm$^{-2}$ on each side). Because the perturbed dOH features overlap with the bulk OH stretching bands in the IP and OP spectra (Fig. 3B), we considered three baselines shown as the dashed horizontal gray lines in Fig. 3C for calculating the IP and OP band areas of the perturbed OH feature (Table A1 in the Appendix).

## 3. Astrophysical Implications

Band strength values for water are summarized in Table 1, and the strengths of the three dangling OH features (2dOH, 3dOH, and perturbed dOH) had values (4.6 × 10$^{-18}$, 9.1 × 10$^{-18}$, and 1.8 × 10$^{-17}$ cm molecule$^{-1}$, respectively) an order of magnitude or more lower values those of bulk water molecules in ices and liquid water (1.5–2.3 × 10$^{-16}$ cm molecule$^{-1}$) (Mastrapa et al. 2009; Max & Chapados 2009; Schaaf & Williams 1973). Thus, hydrogen-bonding networks in bulk ice and liquid water markedly enhance IR absorption (Gorai et al. 2020; Ohno et al. 2005; Van Thiel et al. 1957).

Band strengths for the antisymmetric OH stretching vibration of $H_2O$ monomers confined in solid CO, $N_2$, and $O_2$ matrices (Ehrenfreund et al. 1996) are also shown in Table 1, ranging from $3.3 \times 10^{-18}$ to $1.1 \times 10^{-17}$ cm molecule$^{-1}$, depending on the matrix. These values are similar to those of 20 K dangling OH features determined here. Although dangling OH bonds have been assumed to have a similar band strength to bulk water ice (McCoustra & Williams 1996), the present findings suggest that IR band strengths of dangling OH bonds have monomer-like rather than bulk-water properties.

Our experimental results also indicate that band strengths are ordered as 2dOH < 3dOH < perturbed-dOH as the peak wavenumbers decrease (3720 cm$^{-1}$ > 3697 cm$^{-1}$ > 3650–3640 cm$^{-1}$, respectively) (Fig. 2 and 3 and Table 1). The higher band strength value of 3dOH compared with 2dOH suggests that the dipole moment of a three-coordinated molecule at the amorphous water surface is enhanced by surrounding water molecules (Gregory et al. 1997; Hodgson & Haq 2009). The increase in band strength of the perturbed-dOH feature may be explained by the creation of hydrogen bonding of 2dOH and 3dOH with CO. Previous experimental and theoretical studies have also reported an increase in IR intensity of dangling OH bonds upon $N_2$ and $CH_4$ adsorption (Hujo et al. 2011; Maté et al. 2021). Our experimentally determined band strengths of 3dOH at 3730–3670 cm$^{-1}$ ($9.1 \times 10^{-18}$ cm molecule$^{-1}$) and perturbed dOH at 3680–3620 cm$^{-1}$ ($1.8 \times 10^{-17}$ cm molecule$^{-1}$) are consistent with theoretical values for free dangling OH at 3700–3640 cm$^{-1}$ ($1.2 \times 10^{-17}$ cm molecule$^{-1}$) and $CH_4$-adsorbed dangling OH at 3670–3610 cm$^{-1}$ ($3.1 \times 10^{-17}$ cm molecule$^{-1}$), respectively (Maté et al. 2021). Therefore, although the blue-shifted dOH‒CO features at around 2151 cm$^{-1}$ have yet to be observed in IR spectra of interstellar ices (He et al. 2022b; McClure et al. 2023), our experimentally determined value for perturbed dOH with CO deposition can be used as a reasonable estimate for the band strength of perturbed-dOH features with other molecules, such as $CH_4$. Using the JWST, McClure et al. (2023) reported a potential perturbed dangling OH feature at 3664 $cm^{-1}$ for icy dust grains in molecular clouds around a Class 0 protostar (McClure et al. 2023). Therefore, our results can contribute to interpretation of JWST and other IR observational data in deriving column densities (or their upper limits) for dangling OH bonds in interstellar ices.

This work was supported by JSPS Kakenhi grant Numbers 24H00264, 23H03987 and 21H01143, the Kurita Water and Environment Foundation (Number 23D004), and the Sumitomo Foundation fiscal 2023 Grant for Basic Science Research Projects (Number 2300811).

**Fig. 1.** Schematics of the IR–MAIRS, normal-incidence IR transmission spectroscopy, and IR reflection–absorption spectroscopy systems. (A) Overview of the IR–MAIRS measurement. Seven oblique-incidence IR transmission measurements were performed for thin samples of thickness $d$ at fixed angles of incidence of $\theta = 45°$ at seven different polarization angles from $\phi = 0°$ (s-polarization) to $90°$ (p-polarization) in $15°$ steps. The solid double-headed arrow indicates the direction of the electric field vector for p-polarized radiation, which can be divided into $x$- and $z$-direction components (dashed double-headed arrows); $\otimes$ indicates the direction of the electric field vector for s-polarized radiation (i.e., the $y$-direction). The IP and OP components of single-beam spectra were calculated using classical least-squares regression (see Appendix for details). (B) Normal-incidence IR transmission spectroscopy at $\theta = 0°$ using an IR-transparent substrate. Because the electric field of the IR light is parallel to the substrate, only IP vibration is observed in the spectrum, and the band shape is given by the transverse optic energy-loss function. (C) IR reflection–absorption spectroscopy with a metallic substrate. Because the electric field of the IR light is perpendicular to the metallic substrate, only OP vibration is observed in the spectrum, and the band shape is given by the longitudinal optic energy-loss function.

**Fig. 2.** IR–MAIRS IP and OP spectra of porous amorphous water at 20 K. (A) Spectra over the range of $4000$–$800$ cm$^{-1}$. (B) Enlarged spectra showing 2dOH and 3dOH features at $3740$–$3670$ cm$^{-1}$. (C), (D) Gaussian fitting (gray line) of 2dOH and 3dOH features for (C) IP and (D) OP spectra. Experimental data (open circles) were reproduced by two Gaussian components for 2dOH (green dashed line) and 3dOH (brown dashed line). (C) For the IP

spectrum, the peak wavenumber, FWHM, and area for the 2dOH feature were $3719.7 \pm 0.3$, $9.8 \pm 0.5$, and $9.5 \pm 0.9 \times 10^{-4}$ cm$^{-1}$, and for the 3dOH feature, they were $3696.9 \pm 0.2$, $18.1 \pm 0.6$, and $3.6 \pm 0.2 \times 10^{-3}$ cm$^{-1}$, respectively. (D) For the OP spectrum, the peak wavenumber, FWHM, and area for the 2dOH feature were $3719.6 \pm 0.3$, $9.5 \pm 0.6$, and $7.3 \pm 1.6 \times 10^{-4}$ cm$^{-1}$, and for the 3dOH feature, they were $3696.9 \pm 0.4$, $17.0 \pm 2.0$, and $2.8 \pm 0.4 \times 10^{-3}$ cm$^{-1}$, respectively. Error bars were obtained from five independent measurements. Porous amorphous water was prepared at 20 K with 32 min of water exposure (H$_2$O with 3.5 mol.% HDO) at $5.6 \times 10^{-6}$ Pa.

**Fig. 3.** IP and OP spectra of porous amorphous water at 20 K before and after CO deposition. (A), (B) IP and OP spectra at (A) 3740–3670 and (B) 3800–3600 cm$^{-1}$ for dangling OH features (a) before CO deposition, (b) after CO deposition for 40 min at $7.1 \times 10^{-7}$ Pa, and (c) differential IP and OP spectra ((b) − (a)). (C) Enlarged differential IP and OP spectra of (c) in the panel B. The dashed horizontal gray lines in differential IP and OP spectra show the baselines for calculation of the band areas of the perturbed dOH features (Table A1 in the Appendix). (D) IP and OP spectra at 2170–2120 cm$^{-1}$ for the stretching band of CO (a) before and (b) after CO deposition for 40 min at $7.1 \times 10^{-7}$ Pa. (E), (F) Gaussian fitting (gray line) of the CO stretching band for (E) IP and (F) OP spectra. Experimental data (open circles) were reproduced using two Gaussian components for dOH–CO (green dashed line) and other-CO (brown dashed line). (E) For the IP spectrum, the peak wavenumber, FWHM, and area for the dOH–CO feature were $2150.8 \pm 0.1$, $11.2 \pm 0.1$, and $7.8 \pm 0.2 \times 10^{-3}$ cm$^{-1}$, and those for the other-CO feature were $2138.4 \pm 0.1$, 8.9

± 0.1, and 1.7 ± 0.1 × 10$^{-2}$ cm$^{-1}$, respectively. (F) For the OP spectrum, the peak wavenumber, FWHM, and area for the dOH–CO feature were 2150.5 ± 0.1, 11.9 ± 0.2, and 5.6 ± 0.2 × 10$^{-3}$ cm$^{-1}$, and those for the other-CO feature were 2138.8 ± 0.1, 9.0 ± 0.1, and 1.1 ± 0.1 × 10$^{-2}$ cm$^{-1}$, respectively. Error bars were obtained from five independent measurements. Porous amorphous water was prepared at 20 K with 32 min of water exposure (H$_2$O with 3.5 mol.% HDO) at 5.6 × 10$^{-6}$ Pa, before CO exposure.

**Fig. 4.** IP and OP spectra of nonporous amorphous water at 20 K before and after CO deposition. (A) IP and OP spectra at 3740–3670 cm$^{-1}$ for the 2dOH and 3dOH features (a) at 90 K, (b) at 20 K before CO deposition, and (c) after CO deposition at 20 K for 12 min at 1.6 × 10$^{-7}$ Pa. (B) IP and OP spectra at 2170–2120 cm$^{-1}$ for the stretching band of CO (a) at 90 K, (b) at 20 K before CO deposition, and (c) after CO deposition at 20 K for 12 min at 1.6 × 10$^{-7}$ Pa. (C), (D) Gaussian fitting (gray line) of the CO stretching band for (C) IP and (D) OP spectra. Experimental data (open circles) were reproduced with two Gaussian components for dOH-CO (green dashed line) and other-CO (brown dashed line). (C) For the IP spectrum, the peak wavenumber, FWHM, and area for the dOH-CO feature were 2148.7 ± 1.0, 12.0 ± 1.6, and 7.3 ± 0.9 × 10$^{-4}$ cm$^{-1}$, and those for the other-CO feature were 2139.4 ± 0.3, 8.5 ± 0.4, and 9.5 ± 2.4 × 10$^{-4}$ cm$^{-1}$, respectively. (D) For the OP spectrum, the peak wavenumber, FWHM, and area for the dOH-CO feature were 2154.3 ± 0.4, 10.8 ± 1.4, and 4.9 ± 0.6 × 10$^{-4}$ cm$^{-1}$, and those for the other-CO feature were 2139.9 ± 0.1, 8.8 ± 0.5, and 8.5 ± 1.2 × 10$^{-4}$ cm$^{-1}$, respectively. Error bars are obtained from five

independent measurements. Nonporous amorphous water was prepared at 90 K with 32 min of water exposure (H$_2$O with 3.5 mol.% HDO) at $5.6 \times 10^{-6}$ Pa. (E) Schematic of possible configurations of a CO molecule adsorbed on a three-coordinated dangling OH bond on the surface of nonporous amorphous water at 20 K.

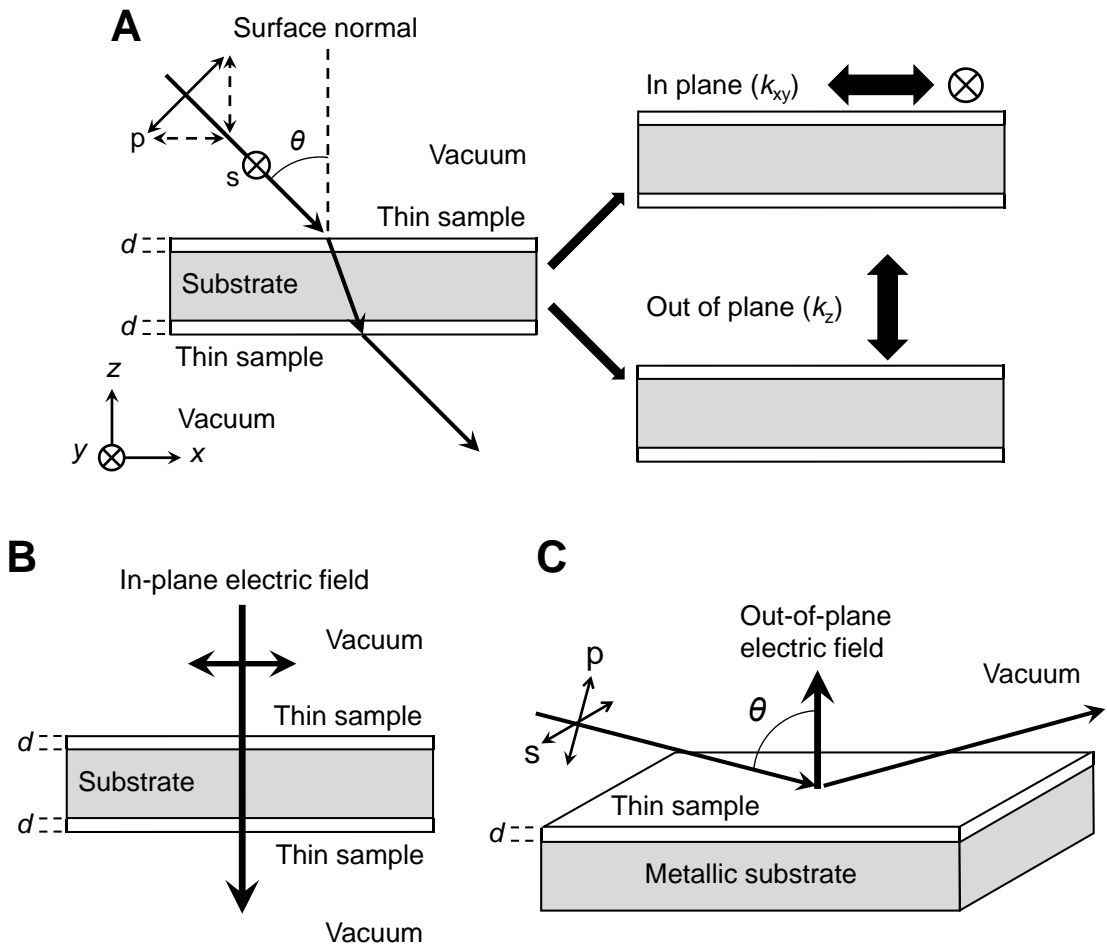

**Fig. 1.**

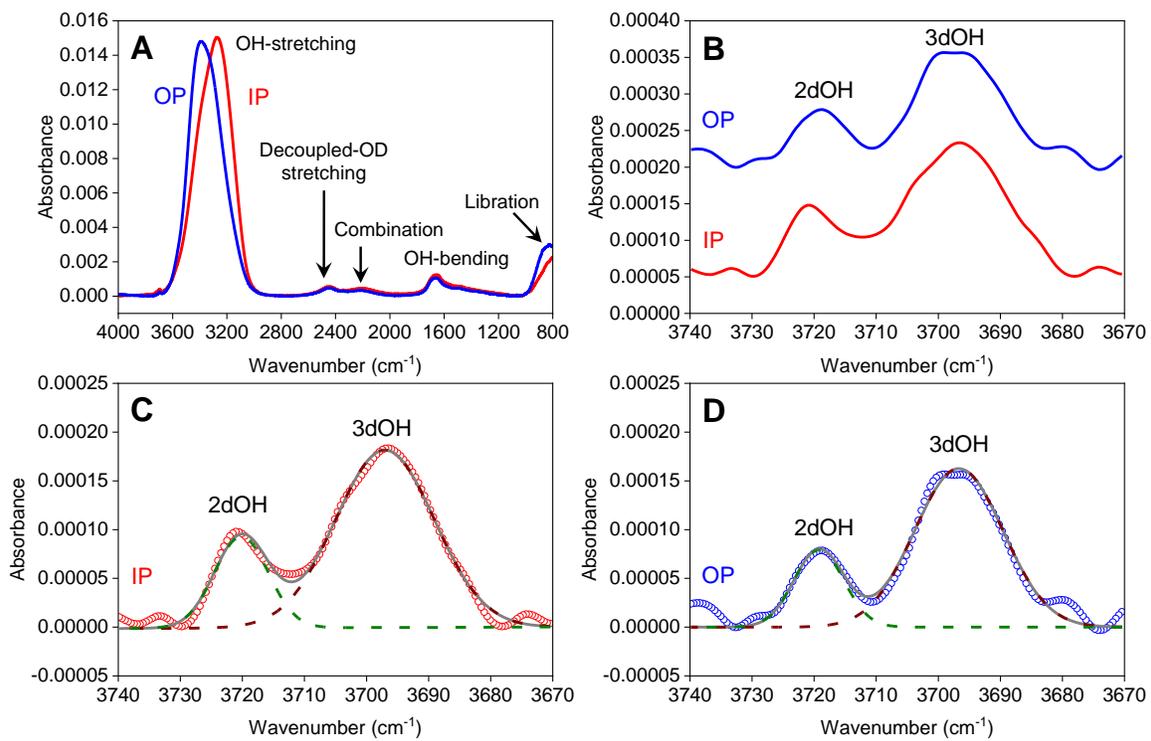

**Fig. 2.**
27

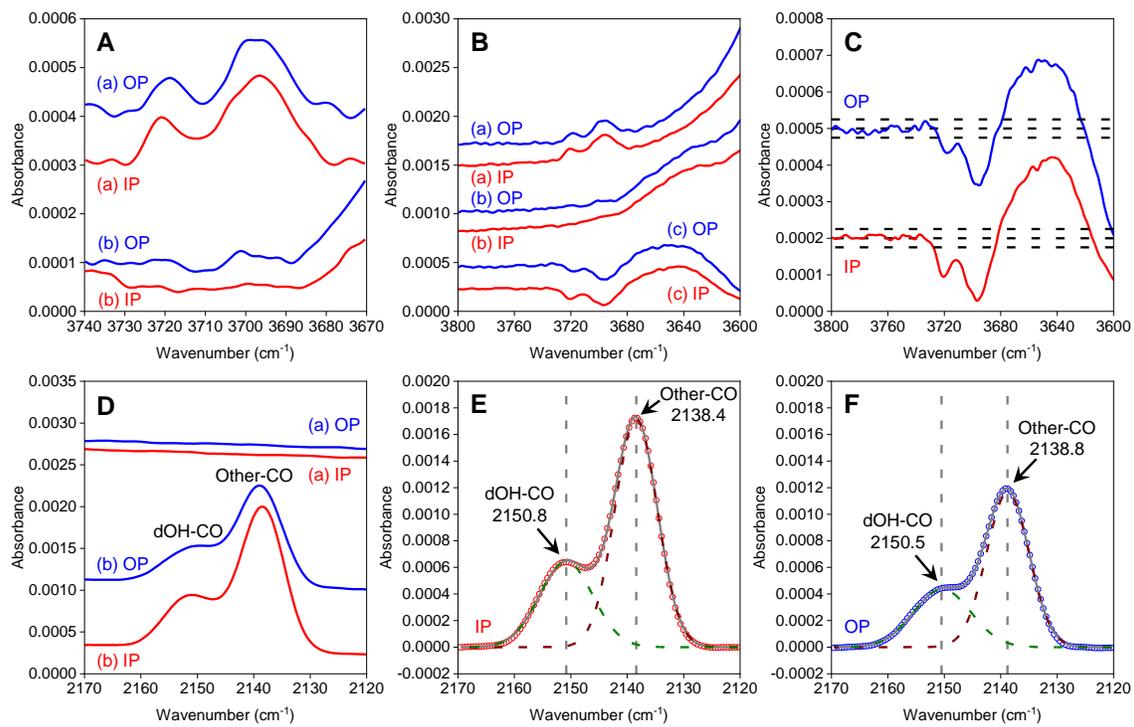

**Fig. 3.**

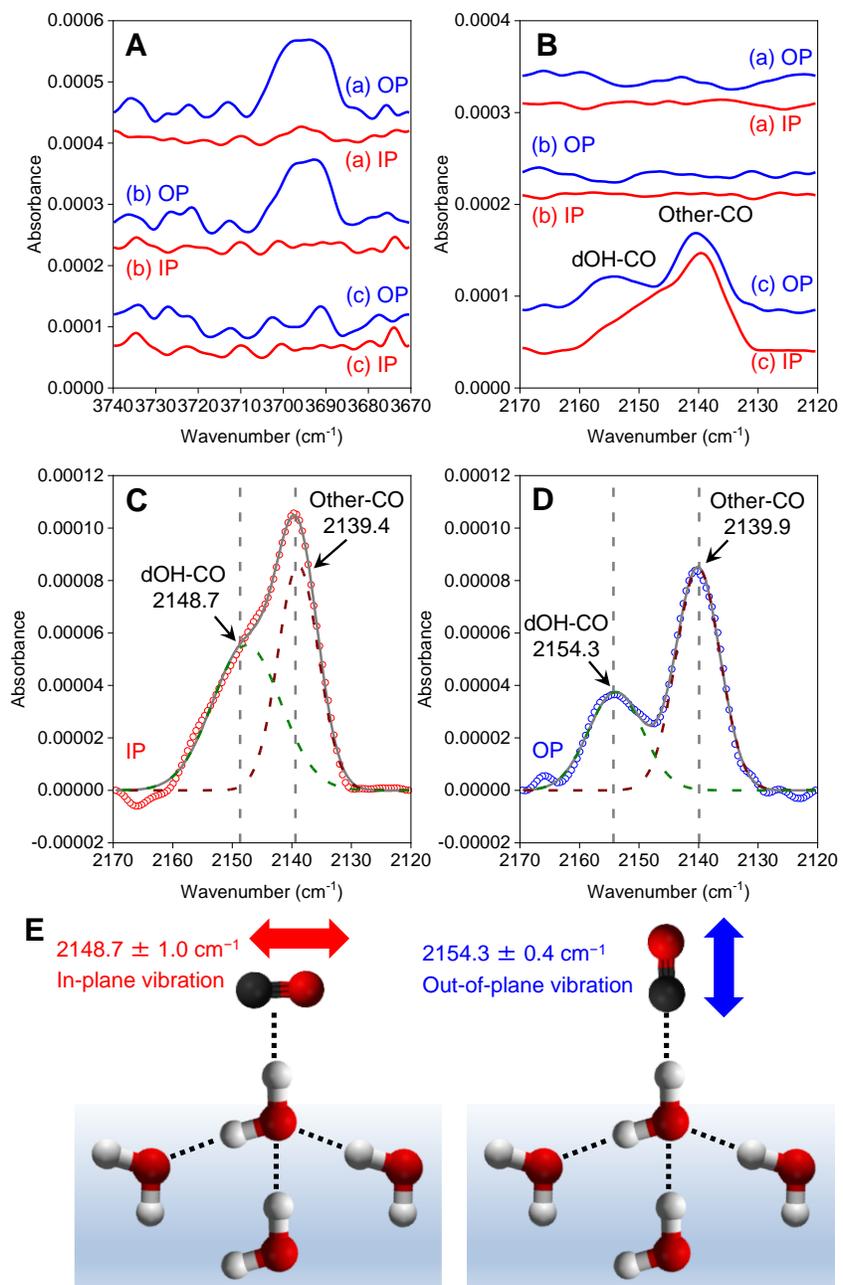

**Fig. 4.**

Table 1. Summary of band strengths (cm molecule$^{-1}$) for water

| Species | Band strength (wavenumber range) |
|---|---|
| Two-coordinated dangling OH at 20 K (this study)[a] | $4.6 \pm 1.6 \times 10^{-18}$ (3740–3700) |
| Three-coordinated dangling OH at 20 K (this study)[b] | $9.1 \pm 1.0 \times 10^{-18}$ (3730–3670) |
| Perturbed dangling OH with CO at 20 K (this study)[a] | $1.8 \pm 0.1 \times 10^{-17}$ (3680–3620) |
| Three-coordinated dangling OH at 90 K (Nagasawa et al. 2021, 2022) | $1.4 \pm 0.3 \times 10^{-17}$ (3710–3680) |
| Dangling OH (calculation) (Maté et al. 2021) | $1.2 \times 10^{-17}$ (3700–3640) |
| Perturbed dangling OH with $CH_4$ (calculation) (Maté et al. 2021) | $3.1 \times 10^{-17}$ (3670–3610) |
| Amorphous water at 15 K (Mastrapa et al. 2009) | $1.9 \times 10^{-16}$ (4320–2933) |
| Crystalline ice at 267–273 K[c] (Schaaf & Williams 1973) | $2.3 \times 10^{-16}$ (3900–2700) |
| Liquid water at 295–298 K[c] (Max & Chapados 2009) | $1.5 \times 10^{-16}$ (3900–2700) |
| Monomer in solid CO at 10 K (Ehrenfreund et al. 1996) | $1.1 \times 10^{-17}$ (3707)[d] |
| Monomer in solid $N_2$ at 10 K (Ehrenfreund et al. 1996) | $7.8 \times 10^{-18}$ (3726)[d] |
| Monomer in solid $O_2$ at 10 K (Ehrenfreund et al. 1996) | $3.3 \times 10^{-18}$ (3732)[d] |

[a] Porous amorphous water was prepared at 20 K with 32 min of water exposure ($H_2O$ with 3.5 mol.% HDO) at $5.6 \times 10^{-6}$ Pa.

[b] Nonporous amorphous water was prepared at 90 K with 32 min of water exposure ($H_2O$ with 3.5 mol.% HDO) at $5.6 \times 10^{-6}$ Pa, and cooled down to 20 K.

[c] See Nagasawa et al. (2021) for details of the band strength calculations for crystalline ice and liquid water.

[d] Asymmetric OH stretching vibration

# APPENDIX

## A.1. Analytical expressions for the absorbance of a thin sample

For reference, the analytical expressions for the absorbance of a thin sample measured by normal-incidence transmission and reflection–absorption using a metallic substrate are summarized below. There are elegantly written textbooks as well as articles on the electromagnetic theoretical treatment of IR spectroscopy (Hasegawa 2017; Itoh et al. 2009; Tolstoy et al. 2003; Yamamoto & Ishida 1994; Yamamoto & Masui 1996). The key approximations to derive analytical expressions for the absorbance of a thin sample are the thin-film approximation (eq. A1) and the weak-absorption approximation (eq. A2):

$$\frac{d}{\lambda} \ll 1 \quad (A1)$$

$$\frac{\Delta S}{S^b} \ll 1 \quad (A2)$$

where $\Delta S = S^b - S^s$. When these two approximations are satisfied, the analytical expression for the thin sample's absorbance is largely simplified (Hasegawa 2017; Itoh et al. 2009).

As described in the section 2.1. in the main text, the absorbance of a thin sample ($A_{\text{thin}}^{\theta=0°}$) for normal-incidence measurement of a typical five-layer, symmetric double-sided sample system (Fig. 1B) is expressed with equations (A3)-(A5):(Itoh et al. 2009)

$$A_{\text{thin}}^{\theta=0°} = -\log_{10}\frac{S^s}{S^b} = \frac{8\pi da}{\lambda \ln 10} f_{\text{TO}} \quad (A3)$$

$$a = \frac{1}{n_v + n_s} + \left(\frac{n_v - n_s}{n_v + n_s}\right)^4 \left\{1 - \left(\frac{n_v - n_s}{n_v + n_s}\right)^4\right\}^{-1} \left(\frac{2n_s}{n_s^2 - n_v^2}\right) \quad (A4)$$

$$f_{\text{TO}} = \text{Im}[\overline{\varepsilon_{xy}}] = \text{Im}[(n_{xy} + ik_{xy})^2] = 2n_{xy}k_{xy} \quad (A5)$$

$f_{\text{TO}}$ is expressed using the *x*- and *y*-components (surface-parallel components) of the complex permittivity ($\overline{\varepsilon_{xy}}$) or complex refractive index ($n_{xy} + ik_{xy}$) of the thin sample, where $\overline{\varepsilon} = (n + ik)^2$. A uniaxial system ($k_x = k_y = k_{xy}$) is assumed for simplicity in this study. Equations (A3)-(A5) show that $A_{\text{thin}}^{\theta=0°}$ is influenced by $n_v$ and $n_s$, despite $A_{\text{thin}}^{\theta=0°}$ being obtained by calculating the ratio between $S^s$ (sample and substrate) and $S^b$ (substrate only). Furthermore, only surface-parallel (IP) vibration is observed in the IR spectrum of the thin sample, whose band shape is expressed as $f_{\text{TO}}$ (Hasegawa 2017; Hasegawa & Shioya 2020; Itoh et al. 2009; Umemura et al. 1990).

For reflection–absorption measurements (Fig. 1C), both the real and imaginary parts of the complex permittivity of the metal substrate ($\overline{\varepsilon}_s$) are significantly larger in the IR range than those of the thin sample ($\overline{\varepsilon}$) and vacuum ($\varepsilon_v$, real value); i.e., $|\varepsilon_s| \gg |\varepsilon|$ and $|\varepsilon_s| \gg \varepsilon_v$. Reflection–absorbance using p-polarized IR light ($A_p^{\text{RA}}$) can be expressed using equations A6 and A7, when $|\varepsilon_s| \gg \varepsilon_v \tan^2\theta$ is satisfied; i.e., the angle of incidence $\theta$ is less than approximately 85°:(Hasegawa 2017; Yamamoto & Ishida 1994)

$$A_p^{\text{RA}} = -\log_{10}\frac{S^s}{S^b} = \frac{8\pi d}{\lambda \ln 10}\frac{n_v^3 \sin^2\theta}{\cos\theta} f_{\text{LO}} \quad (A6)$$

$$f_{\text{LO}} = \text{Im}\left[-\frac{1}{\overline{\varepsilon_z}}\right] = \text{Im}\left[\frac{1}{(n_z + ik_z)^2}\right] = \frac{2n_z k_z}{(n_z^2 + k_z^2)^2} \quad (A7)$$

$f_{\text{LO}}$ is expressed using the *z*-component (surface-perpendicular component) of the complex permittivity ($\overline{\varepsilon_z}$) or complex refractive index ($n_z + ik_z$) of the thin sample. Reflection–absorption measurements have the electric field of the IR light perpendicular to the metallic substrate; therefore, only the surface-perpendicular (OP) vibration is observed in the IR spectrum of the thin sample, and the band shape is given by $f_{\text{LO}}$ .(Hasegawa 2017; Hasegawa & Shioya 2020; Umemura et al. 1990) Negligible reflection–absorbance is induced by s-polarized IR light (Hasegawa 2017).

**A.2. Low-temperature, ultrahigh-vacuum IR–MAIRS apparatus**

Our previous articles detail the measurement and analytical procedures of low-temperature, ultrahigh-vacuum IR–MAIRS (Hama et al. 2020; Nagasawa et al. 2021, 2022), so only a brief outline is given here. The apparatus comprises a vacuum chamber and Fourier transform infrared (FTIR) spectrometer. The chamber is evacuated to ultrahigh-vacuum conditions (base pressure $10^{-7}$ Pa at room temperature and $10^{-8}$ Pa at 20 K) using a turbo molecular pump. A Si(111) substrate (40 × 40 × 1 mm) and a copper sample holder are connected with indium solder. The sample holder is connected to the cold head of a closed-cycle helium refrigerator and installed in the vacuum chamber using a bore-through rotary feedthrough. Because the refrigerator and substrate are freely rotatable via the rotary feedthrough, the angle of incidence of the IR beam ($\theta$) can be varied by rotating the substrate, with an accuracy of ±0.5°. The substrate (in the chamber) is installed in the sample compartment of the FTIR spectrometer in transmission geometry across two ZnSe windows in the chamber. The temperature of the substrate is measured using a Si diode sensor placed in the sample holder, and controlled with an accuracy of ±0.2 K.

In our experiments, purified $H_2O$ (resistivity ≥18.2 MΩ cm at 298 K) from a Millipore Milli-Q water purification system was mixed with 2.0 wt.% (1.8 mol.%) $D_2O$ (deuteration degree > 99.9%; Merck, USA) to obtain a water sample containing about 3.5 mol.% HDO with a negligible amount of $D_2O$. The water was first degassed by several freeze–pump–thaw cycles. The CO deposition experiments used CO gas (>99.95% purity; Kotobuki Sangyo Co. Ltd., Japan). Gas pressure was measured with a cold-cathode gauge, applying the gas correction factors for $H_2O$ (1.25 ± 0.44) and CO (1.02 ± 0.08) (Nakao 1975). Amorphous water was produced on the Si substrate at 20 K by background water deposition at $5.6 \pm 2.0 \times 10^{-6}$ Pa. The water flux was estimated as $2.0 \pm 0.7 \times 10^{13}$ molecules $cm^{-2}$ $s^{-1}$. The typical deposition time is 1920 s (32 min), which corresponding to an exposure of $3.9 \pm 1.4 \times 10^{16}$ molecules $cm^{-2}$. We also calculated the column density of water as $4.2 \pm 0.1 \times 10^{16}$ molecules $cm^{-2}$ from the OH stretching band area at 3825–2727 $cm^{-1}$ in the IP and OP spectra using the band strength of $1.9 \times 10^{-16}$ cm $molecule^{-1}$ (Mastrapa et al. 2009), which is in good agreement with the amount of water exposure within the error margins. For reference, the lattice parameters of hexagonal ice indicated ~$1.0 \times 10^{15}$ molecules $cm^{-2}$ on the surface (Petrenko & Whitworth 1999).

**A.3. IR–MAIRS measurements and analysis**

For IR–MAIRS measurements, FT-modulated IR light was passed through an angle-controllable wire-grid linear ZnSe polarizer incorporated in the FTIR spectrometer, and oblique-incidence transmission measurements were taken at $\theta = 45°$ at seven polarization angles from $\phi = 0°$ (s-polarization) to 90° (p-polarization) in 15° steps. A mercury cadmium telluride IR detector was used, and the intensity of the transmitted IR light measured as a single-beam measurement by the FTIR system. The accumulation number of the single-beam measurements was 1000 (615 s) for each polarization angle, with a resolution of 4 cm$^{-1}$. The total measurement time was 4305 s (72 min) for one IR–MAIRS measurement.

The intensity of transmitted IR light, measured as single-beam spectra ($S_{\text{obs}}$) with polarization angle $\phi$ and incident angle $\theta$ (fixed at 45°), was expressed as a linear combination of the IP and OP polarization components ($S_{\text{IP}}$ and $S_{\text{OP}}$, respectively) with weighting ratios ($r_{\text{IP}}$ and $r_{\text{OP}}$, respectively) and nonlinear noise factors $U$ (e.g., reflected IR light).

$$S_{\text{obs}} = r_{\text{IP}} S_{\text{IP}} + r_{\text{OP}} S_{\text{OP}} + U \qquad (A1)$$

Collection of the $j$th single-beam spectrum, $S_{\text{obs},j}$ ($j = 1, 2, …, 7$), at a polarization angle of $\phi_j$ ($j = 1, 2, …, 7$), forms the matrix, $S$, with the linear combination part described using classical least-squares regression as (Equation (A2))

$$S = \begin{pmatrix} S_{\text{obs},1} \\ S_{\text{obs},2} \\ \vdots \\ S_{\text{obs},7} \end{pmatrix} = \begin{pmatrix} r_{\text{IP},1} & r_{\text{OP},1} \\ r_{\text{IP},2} & r_{\text{OP},2} \\ \vdots & \vdots \\ r_{\text{IP},7} & r_{\text{OP},7} \end{pmatrix} \begin{pmatrix} S_{\text{IP}} \\ S_{\text{OP}} \end{pmatrix} + U \equiv R \begin{pmatrix} S_{\text{IP}} \\ S_{\text{OP}} \end{pmatrix} + U \qquad (A2)$$

where $R$ is a matrix of weighting coefficients of $r_{\text{IP},j}$ ($j = 1, 2…7$) and $r_{\text{OP},j}$ ($j = 1, 2…7$) for $S_{\text{IP}}$ and $S_{\text{OP}}$, respectively. The weighting factor matrix, $R$, is expressed as

$$R = \begin{pmatrix} \gamma\cos^2\phi_1 + \sin^2\phi_1(\sin^2\theta\tan^2\theta + \cos^2\theta) & \sin^2\phi_1\tan^2\theta \\ \gamma\cos^2\phi_2 + \sin^2\phi_2(\sin^2\theta\tan^2\theta + \cos^2\theta) & \sin^2\phi_2\tan^2\theta \\ \vdots & \vdots \\ \gamma\cos^2\phi_7 + \sin^2\phi_7(\sin^2\theta\tan^2\theta + \cos^2\theta) & \sin^2\phi_7\tan^2\theta \end{pmatrix} \quad (A3)$$

where $\gamma$ is the intensity ratio of s-polarized to p-polarized light when measured without a substrate (Itoh et al. 2009; Nagasawa et al. 2022; Shioya et al. 2019). $U$ is a 'garbage matrix' that received the non-linear responses to $R$, which means that noise factors were rejected in the classical least-squares regression calculation and pooled in $U$ as an error term. Equation (A2) allowed $S_{IP}$ and $S_{OP}$ to be calculated as the least-squares solution of the classical least-squares regression equation as

$$\begin{pmatrix} S_{IP} \\ S_{OP} \end{pmatrix} = (R^T R)^{-1} R^T S \quad (A4)$$

where $S_{IP}$ and $S_{OP}$ correspond to IP and OP single-beam spectra, respectively. Hence, the IP and OP absorbance spectra ($A_{IP}$ and $A_{OP}$, respectively) were obtained as

$$A_{IP} = -\log_{10} \frac{S_{IP}^s}{S_{IP}^b} \quad (A5)$$

$$A_{OP} = -\log_{10} \frac{S_{OP}^s}{S_{OP}^b} \quad (A6)$$

where the superscripts $b$ and $s$ indicate background and sample measurements, respectively. Baselines of the IP and OP absorbance spectra in the figures in this study were corrected and offset for clarity. IP and OP spectral vertical axes (absorbance) in the Figures are $A_{IP}$ and $n^4 H A_{OP}$ from Equations (11) and (12), respectively, reflecting $k_{xy}$ and $k_z$ with the common ordinate scale as $A_{IP}/n^4 H A_{OP} = k_{xy}/k_z$. The obtained experimental spectra were analyzed with Origin Pro 2021b software (OriginLab Corp., USA), and the results of Gaussian fitting of the experimental spectra are summarized in Table A1-A5.

**A.4. CO deposition on porous amorphous water at 20 K**

Figure 5 shows typical IP and OP spectra of porous amorphous water at 20 K as a function of CO deposition time at $7.1 \times 10^{-7}$ Pa. The 2dOH and 3dOH features decrease after 20 minutes of deposition, and completely vanishes after 40-45 minutes (Fig. 5A and B). Regarding the CO stretching bands (Fig. 5C and D), the band area of the blue-shifted dOH-CO feature saturated and almost constant after CO deposition for 40-45 minutes, whereas that of the red-shifted other-CO feature continues to increase (Table A6). These results support that the blue- and red-shifted features are associated with CO molecules interacting with dangling OH and bonded-OH (or other non-dangling OH sites), respectively.

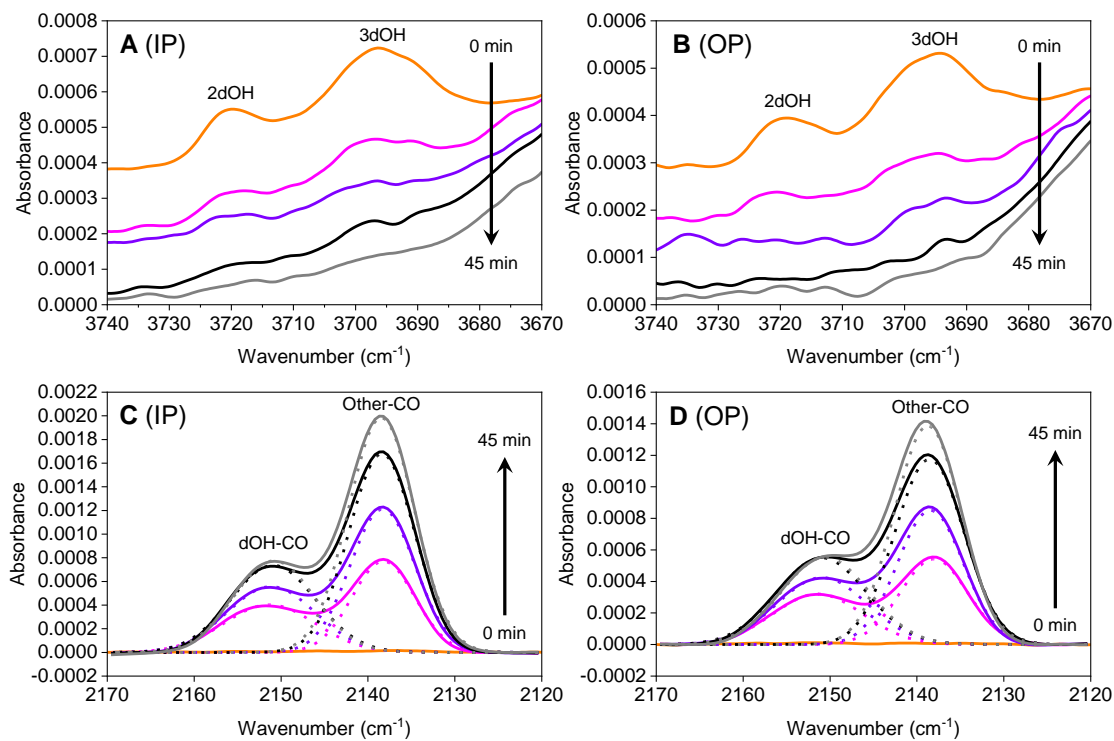

**Fig. 5.** Typical IP and OP spectra of porous amorphous water at 20 K before CO deposition (0 min, orange) and after CO deposition at $7.1 \times 10^{-7}$ Pa for 20 min (pink), 30 min (violet), 40 min (black), and 45 min (gray). (A) IP and (B) OP spectra at 3740–3670 cm$^{-1}$ for dangling OH features. (C) IP and (D) OP spectra at 2170–2120 cm$^{-1}$ for the stretching band of CO. The Gaussian fits have been overlaid to aid the eye (see also Table A6). Porous amorphous water was prepared at 20 K with 32 min of water exposure (H$_2$O with 3.5 mol.% HDO) at $5.6 \times 10^{-6}$ Pa, before CO exposure.

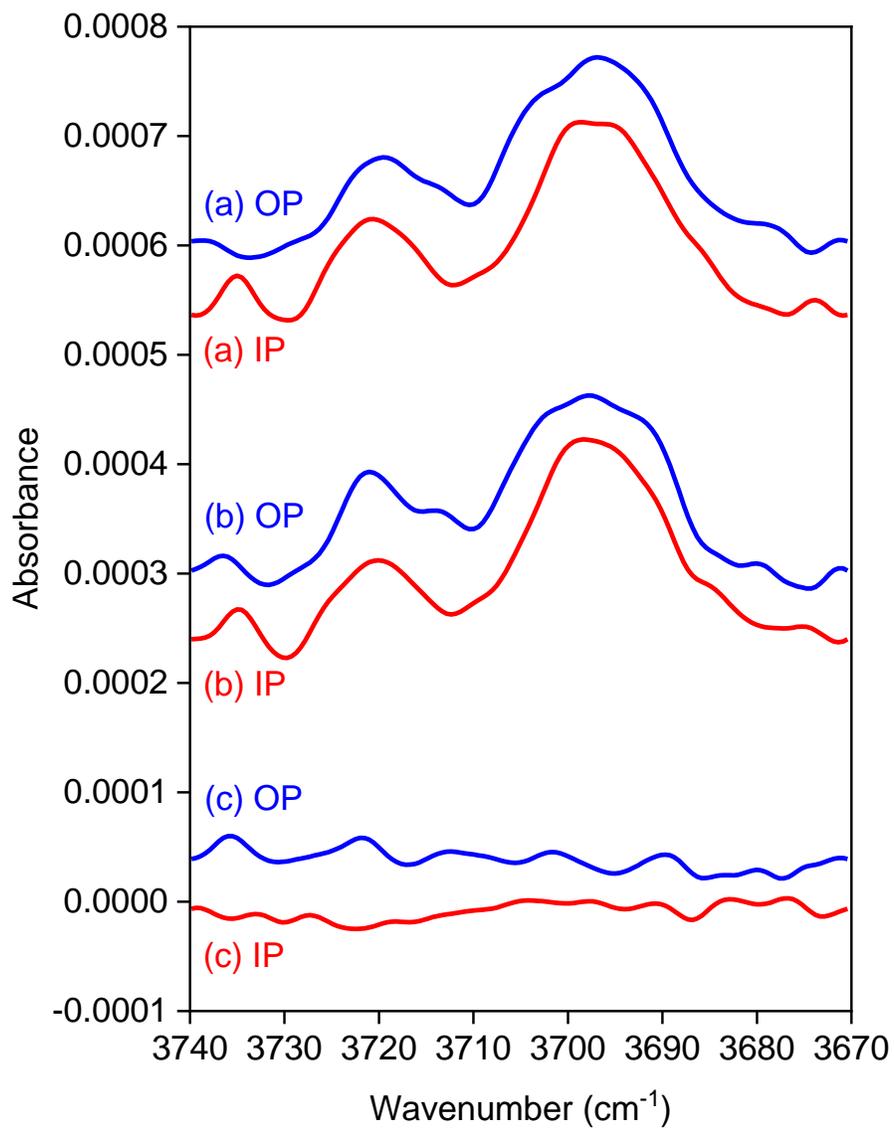

**Fig. 6.** IP and OP spectra of porous amorphous water at 3740–3670 cm$^{-1}$ for the 2dOH and 3dOH features. (a) Porous amorphous water prepared at 20 K with 32 min of water exposure (H$_2$O with 3.5 mol.% HDO) at $5.6 \times 10^{-6}$ Pa. (b) Porous amorphous water left for 88 min at 20 K. (c) differential IP and OP spectra ((b) − (a)).

Table A1. Summary of IP and OP band areas and column densities of the 3dOH and 2dOH features for porous amorphous water at 20 K.[a,b]

| Species (wavenumber range) | IP band area ($cm^{-1}$) | OP band area ($cm^{-1}$) | Column density (molecules $cm^{-2}$) |
|---|---|---|---|
| 2dOH[c] (3740–3700 $cm^{-1}$) | $9.5 \pm 0.9 \times 10^{-4}$ | $7.3 \pm 1.6 \times 10^{-4}$ | $6.5 \pm 2.2 \times 10^{14}$ [e] <br> $[3.2 \pm 1.1 \times 10^{14}]$ [f] |
| 3dOH[c] (3730–3670 $cm^{-1}$) | $3.6 \pm 0.2 \times 10^{-3}$ | $2.8 \pm 0.4 \times 10^{-3}$ | $1.2 \pm 0.1 \times 10^{15}$ [e] <br> $[6.1 \pm 0.3 \times 10^{14}]$ [f] |
| Perturbed dOH[d] (3730–3670 $cm^{-1}$) | $1.0 \pm 0.1 \times 10^{-2}$ | $8.9 \pm 0.6 \times 10^{-3}$ | $1.9 \pm 0.2 \times 10^{15}$ [e] <br> $[9.3 \pm 1.0 \times 10^{14}]$ [f] |

[a] Porous amorphous water was prepared at 20 K with 32 min of water exposure ($H_2O$ with 3.5 mol.% HDO) at $5.6 \times 10^{-6}$ Pa.

[b] The errors reflect the standard deviation in independent statistical experiments.

[c] Before CO deposition.

[d] After CO deposition for 40 min at $7.1 \times 10^{-7}$ Pa.

[e] Column density adsorbed on both sides of the Si substrate.

[f] Column density adsorbed on each side of the Si substrate.

Table A2. Summary of IP and OP band areas and column densities of the dOH-CO and other-CO features for CO deposited on porous amorphous water at 20 K.[a,b]

| Species (wavenumber range) | IP band area (cm$^{-1}$) | OP band area (cm$^{-1}$) | Column density (molecules cm$^{-2}$) |
|---|---|---|---|
| dOH-CO (2170–2130 cm$^{-1}$) | $7.8 \pm 0.2 \times 10^{-3}$ | $5.6 \pm 0.2 \times 10^{-3}$ | $1.9 \pm 0.2 \times 10^{15}$ [c]<br>[$9.3 \pm 1.0 \times 10^{14}$] [d] |
| Other-CO (2160–2120 cm$^{-1}$) | $1.7 \pm 0.1 \times 10^{-2}$ | $1.1 \pm 0.1 \times 10^{-2}$ | $3.9 \pm 0.4 \times 10^{15}$ [c]<br>[$2.0 \pm 0.2 \times 10^{14}$] [d] |
| Total | $2.4 \pm 0.1 \times 10^{-2}$ | $1.7 \pm 0.1 \times 10^{-2}$ | $5.8 \pm 0.6 \times 10^{15}$ [c]<br>[$2.9 \pm 0.3 \times 10^{15}$] [d] |

[a] Porous amorphous water was prepared at 20 K with 32 min of water exposure (H$_2$O with 3.5 mol.% HDO) at $5.6 \times 10^{-6}$ Pa. CO was deposited on porous amorphous water at 20 K for 40 min at $7.1 \times 10^{-7}$ Pa.

[b] The errors reflect the standard deviation in independent statistical experiments.

[c] Column density adsorbed on both sides of the Si substrate.

[d] Column density adsorbed on each side of the Si substrate.

Table A3. Summary of IP and OP band areas and column densities of the 3dOH and 2dOH features for nonporous amorphous water at 20 K before CO deposition.[a,b]

| Species (wavenumber range) | IP band area (cm$^{-1}$) | OP band area (cm$^{-1}$) | Column density (molecules cm$^{-2}$) |
|---|---|---|---|
| 2dOH (3740–3700 cm$^{-1}$) | -[c] | -[c] | -[c] |
| 3dOH (3730–3670 cm$^{-1}$) | -[c] | $1.4 \pm 0.3 \times 10^{-3}$ | $1.6 \pm 0.2 \times 10^{14}$ [d] <br> [$8.2 \pm 0.9 \times 10^{13}$] [e] |

[a] Nonporous amorphous water was prepared at 90 K with 32 min of water exposure (H$_2$O with 3.5 mol.% HDO) at $5.6 \times 10^{-6}$ Pa.

[b] The errors reflect the standard deviation in independent statistical experiments.

[c] Below the limit of detection of IR-MAIRS

[d] Column density adsorbed on both sides of the Si substrate.

[e] Column density adsorbed on each side of the Si substrate.

Table A4. Summary of IP and OP band areas and column densities of the dOH-CO and other-CO features for CO deposited on nonporous amorphous water at 20 K.[a,b]

| Species (wavenumber range) | IP band area (cm$^{-1}$) | OP band area (cm$^{-1}$) | Column density (molecules cm$^{-2}$) |
|---|---|---|---|
| dOH-CO (2170–2130 cm$^{-1}$) | $7.3 \pm 0.9 \times 10^{-4}$ | $4.9 \pm 0.5 \times 10^{-4}$ | $1.6 \pm 0.2 \times 10^{14}$ [c] <br> [$8.2 \pm 0.9 \times 10^{13}$] [d] |
| Other-CO (2160–2120 cm$^{-1}$) | $9.9 \pm 2.4 \times 10^{-4}$ | $8.5 \pm 1.2 \times 10^{-4}$ | $2.4 \pm 0.3 \times 10^{14}$ [c] <br> [$1.2 \pm 0.1 \times 10^{14}$] [d] |
| Total | $1.7 \pm 0.3 \times 10^{-3}$ | $1.3 \pm 0.2 \times 10^{-3}$ | $4.0 \pm 0.5 \times 10^{14}$ [c] <br> [$2.0 \pm 0.2 \times 10^{14}$] [d] |

[a] Nonporous amorphous water was prepared at 90 K with 32 min of water exposure (H$_2$O with 3.5 mol.% HDO) at $5.6 \times 10^{-6}$ Pa. CO was deposited on nonporous amorphous water at 20 K for 12 min at $1.6 \times 10^{-7}$ Pa.

[b] The errors reflect the standard deviation in independent statistical experiments.

[c] Column density adsorbed on both sides of the Si substrate.

[d] Column density adsorbed on each side of the Si substrate.

Table A5. Summary of IP and OP band areas of the 3dOH feature for nonporous amorphous water at 90 and 20 K.[a,b]

| Species (wavenumber range) | IP band area (cm$^{-1}$) | OP band area (cm$^{-1}$) |
| --- | --- | --- |
| 3dOH at 90 K (3730–3670 cm$^{-1}$) | -[c] | $2.0 \pm 0.4 \times 10^{-3}$ |
| 3dOH at 20 K (3730–3670 cm$^{-1}$) | -[c] | $1.4 \pm 0.3 \times 10^{-3}$ |

[a] Nonporous amorphous water was prepared at 90 K with 32 min of water exposure (H$_2$O with 3.5 mol.% HDO) at $5.6 \times 10^{-6}$ Pa.

[b] The errors reflect the standard deviation in independent statistical experiments.

[c] Below the limit of detection of IR-MAIRS.

Table A6. Summary of peak wavenumber, FWHM, and band area of Gaussian fitting for the dOH-CO and other-CO features of CO on porous amorphous water at 20 K.[a]

| | dOH-CO | | | | | |
|---|---|---|---|---|---|---|
| | IP | | | OP | | |
| Deposition time | Peak wavenumber ($cm^{-1}$) | FWHM ($cm^{-1}$) | Area[b] ($cm^{-1}$) | Peak wavenumber ($cm^{-1}$) | FWHM ($cm^{-1}$) | Area[b] ($cm^{-1}$) |
| 20 min | 2151.3 | 12.4 | $5.3 \times 10^{-3}$ | 2151.4 | 12.2 | $4.2 \times 10^{-3}$ |
| 30 min | 2151.1 | 12.0 | $7.1 \times 10^{-3}$ | 2150.9 | 12.3 | $5.6 \times 10^{-3}$ |
| 40 min | 2150.9 | 11.5 | $9.0 \times 10^{-3}$ | 2150.9 | 12.1 | $7.1 \times 10^{-3}$ |
| 45 min | 2150.8 | 11.0 | $9.0 \times 10^{-3}$ | 2150.6 | 11.8 | $7.0 \times 10^{-3}$ |

| | Other CO | | | | | |
|---|---|---|---|---|---|---|
| | IP | | | OP | | |
| Deposition time | Peak wavenumber ($cm^{-1}$) | FWHM ($cm^{-1}$) | Area[b] ($cm^{-1}$) | Peak wavenumber ($cm^{-1}$) | FWHM ($cm^{-1}$) | Area[b] ($cm^{-1}$) |
| 20 min | 2138.2 | 8.6 | $7.0 \times 10^{-3}$ | 2138.0 | 9.2 | $5.3 \times 10^{-3}$ |
| 30 min | 2138.3 | 8.8 | $11.4 \times 10^{-3}$ | 2138.4 | 9.0 | $8.1 \times 10^{-3}$ |
| 40 min | 2138.4 | 8.9 | $15.9 \times 10^{-3}$ | 2138.6 | 9.4 | $11.7 \times 10^{-3}$ |
| 45 min | 2138.5 | 9.0 | $18.9 \times 10^{-3}$ | 2138.8 | 9.0 | $13.4 \times 10^{-3}$ |

[a] The spectral data are shown in Fig. 5. Porous amorphous water was prepared at 20 K with 32 min of water exposure ($H_2O$ with 3.5 mol.% HDO) at $5.6 \times 10^{-6}$ Pa. CO was deposited on porous amorphous water at $7.1 \times 10^{-7}$ Pa.

[b] The standard deviation of the calculated area is less than $1 \times 10^{-4}$ ($cm^{-1}$).